\documentclass{article}
\usepackage{spconf,amsmath,graphicx}
\usepackage{amsmath}
\usepackage{amssymb}
\usepackage{diagbox}
\usepackage{multirow}
\usepackage{booktabs}

\usepackage{physics}
\usepackage{esvect}
\usepackage{acronym}
\usepackage[acronym]{glossaries}

\title{DPT-FSNET: DUAL-PATH TRANSFORMER BASED FULL-BAND AND SUB-BAND FUSION NETWORK FOR SPEECH ENHANCEMENT}
\name{Feng Dang$^{\star \dagger}$ \qquad Hangting Chen$^{\star \dagger}$ \qquad Pengyuan Zhang$^{\star \dagger}$}
  
 \address{$^{\star}$ Key Laboratory of Speech Acoustics \& Content Understanding, Institute of Acoustics, CAS, China \\
          $^{\dagger}$ University of Chinese Academy of Sciences, Beijing, China}
      
\begin{document}
\ninept
\maketitle
\begin{abstract}
Sub-band models have achieved promising results due to their ability to model local patterns in the spectrogram. Some studies further improve the performance by fusing sub-band and full-band information. However, the structure for the full-band and sub-band fusion model was not fully explored. This paper proposes a dual-path transformer-based full-band and sub-band fusion network (DPT-FSNet) for speech enhancement in the frequency domain. The intra and inter parts of the dual-path transformer model sub-band and full-band information, respectively. The features utilized by our proposed method are more interpretable than those utilized by the time-domain dual-path transformer. We conducted experiments on the Voice Bank + DEMAND and Interspeech 2020 Deep Noise Suppression (DNS) datasets to evaluate the proposed method. Experimental results show that the proposed method outperforms the current state-of-the-art.
\end{abstract}

\begin{figure*}[t]
  \centering
  \includegraphics[width=\linewidth,height=0.45\textwidth]{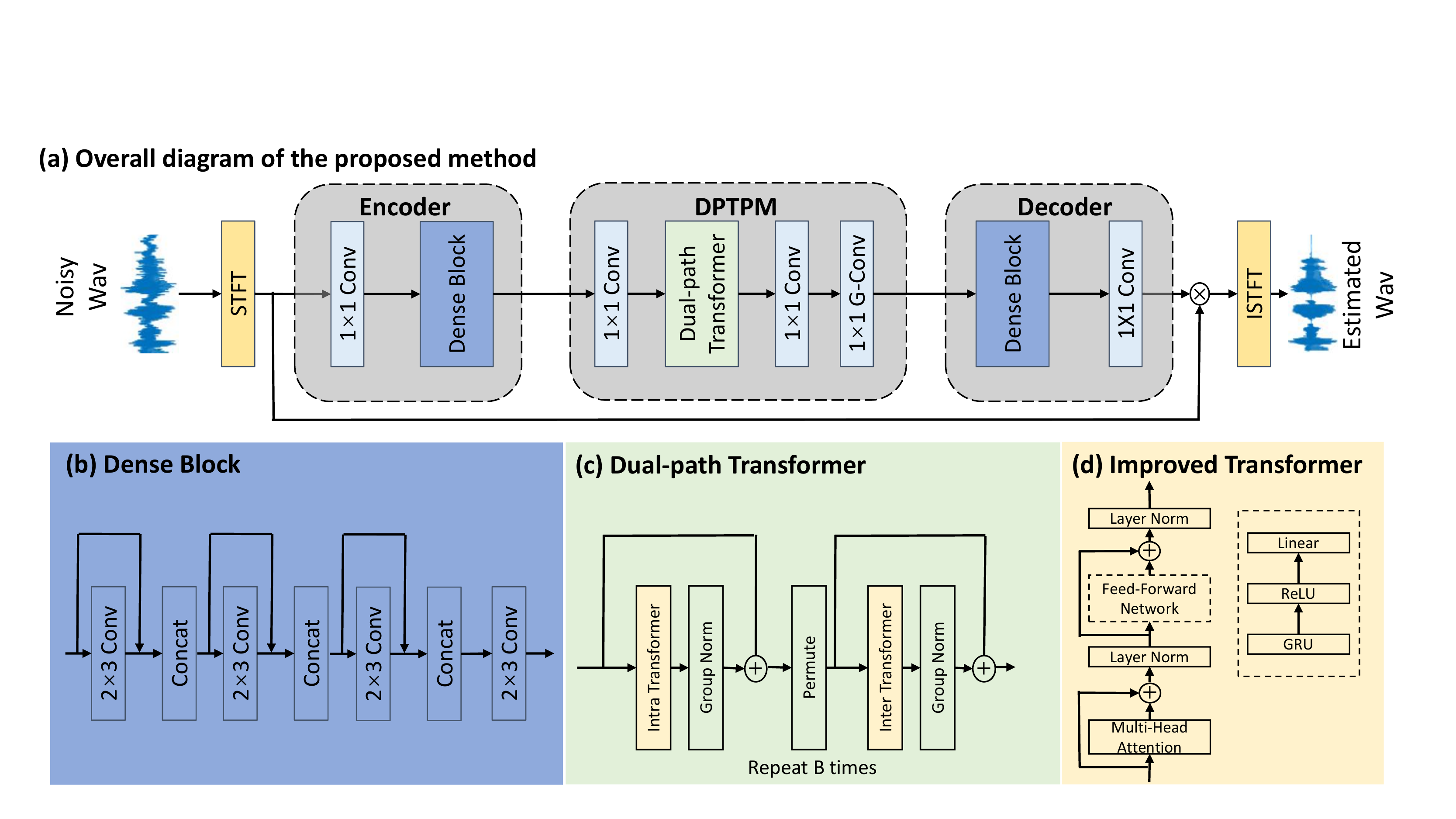}
  \caption{Architecture of the proposed DPT-FSNet. a) The overall diagram of the proposed method b) The detail of the dense block. c) The detail of the dual-path transformer. d) The detail of the improved transformer. }
  \label{fig:speech_production}
\end{figure*}

\section{Introduction}
\label{sec:intro}

Speech enhancement (SE) is a speech processing method that aims to improve the quality and intelligibility of noisy speech by removing noise \cite {loizou2013speech}. It is commonly used as a front-end task for automatic speech recognition, hearing aids, and telecommunications. In recent years, the application of deep neural networks (DNNs) in SE research has received increasing interest.

In general, DNN-based methods can be divided into two major categories: time-domain methods \cite{pandey2019new, defossez20_interspeech,kim2021se} and time-frequency domain (T-F) methods \cite{fu2019metricgan,kim2020t,hao2021fullsubnet,li2021two}. Time-domain methods estimate clean waveforms directly from the noisy raw data in the time domain. Traditional T-F domain methods usually transform the noisy input waveform into a Fourier magnitude spectrum by short-time Fourier transform, modify the spectrum by T-F mask, and reconstruct the enhanced spectrum into enhanced waveform by inverse short-time Fourier transform. They usually use the phase of the noisy mixture, which limits the upper bound of the denoising performance. Recent T-F domain methods using complex spectra as features can preserve phase information and have achieved promising performance \cite{li2021two}.

Sub-band processing is a common method in audio processing \cite{liu20f_interspeech,yang2021multi,lv21_interspeech}, which takes sub-band spectral features as input and output. Previous work \cite{takahashi2017multi} pointed out that the local patterns in the spectrum tend to be different in each frequency band. The sub-band model handles each frequency independently, which allows the sub-band model to focus on the local patterns in the spectrum and therefore achieve good results in SE tasks. In \cite{hao2021fullsubnet} further improves the performance by fusing sub-band and full-band information. 

Recently, dual-path networks \cite{luo2020dual,chen20l_interspeech,zhang2021transmask,wang2021tstnn} have achieved exceptional performance due to their ability to model local and global features of the input sequence. Some studies \cite{chen20l_interspeech,zhang2021transmask} have introduced transformer structures \cite{vaswani2017attention} into dual-path networks, where input elements can interact directly based on self-attention mechanism, to further improve the performance of dual-path networks. However, these studies were based on simple time-domain features and did not further investigate the effect of the input of the dual-path network on the enhancement performance. \cite{wang2021tstnn} did try to change the structure of encoder and decoder to extract more effective inputs for the dual-path network, but still limited to time-domain features.

Inspired by the above problems, we propose a dual-path transformer based full-band and sub-band fusion network (DPT-FSNet) for speech enhancement. Specifically, our proposed model consists of an encoder, decoder, and dual-path transformer. We utilize a convolutional encoder-decoder (CED) structure to extract an efficient latent feature space for the dual-path transformer. Both the encoder and decoder consist of a 1x1 convolutional layer and a dense block \cite{huang2017densely}, where dilated convolutions are utilized inside the dense block for context aggregation. The dual-path transformer is composed of two parts, intra-transformers and inter-transformers. The intra-transformer models sub-band information and the inter-transformer merges the sub-band information from the intra-transformer to model the full-band information. We evaluated our model on the VoiceBank+DEMAND (VCTK+DEMAND) dataset \cite{valentini2016investigating} and  Interspeech 2020 Deep Noise Suppression (DNS) dataset \cite{reddy2020interspeech}. The experimental results show that the proposed model achieves better results than other speech enhancement models.



\section{Improved transformer}
\label{sec:format}

Generally speaking, a transformer consists of an encoder and a decoder \cite{vaswani2017attention}. In this paper, we choose the transformer encoder as our basic block. To avoid confusion, the reference to the transformer in this paper refers to the encoder part of the transformer. The original transformer encoder usually contains three modules: positional encoding, multi-head self-attention, and position-wise feed-forward network. In this paper, our transformer consists of two modules as in \cite{chen20l_interspeech}: multi-head self-attention and modified position-wise feed-forward network.
\subsection{Multi-head self-attention}
\label{sec:majhead}

We used the multi-headed self-attention from \cite{vaswani2017attention}. The multi-headed self-attention module can be formulated as:
\begin{footnotesize}
\begin{equation}
\begin{split}
   Q_{i}=ZW_{i}^{Q},K_{i}=ZW_{i}^{K},V_{i}=ZW_{i}^{V} i \in [1,h]
  \label{eq5}
\end{split}
\end{equation}
\end{footnotesize}
\begin{footnotesize}
\begin{equation}
\begin{split}
   head_{i}&=Attention(Q_{i},K_{i},V_{i})=SoftMax(\frac{Q_{i}K_{i}}{\sqrt{d}})V_{i}
  \label{eq5}
\end{split}
\end{equation}
\end{footnotesize}
\begin{footnotesize}
\begin{equation}
\begin{split}
   MultiHead=Concat(head_{1},\cdots,head_{h})W^{O}
  \label{eq5}
\end{split}
\end{equation}
\end{footnotesize}
\begin{footnotesize}
\begin{equation}
\begin{split}
   Mid = LayerNorm(Z+MultiHead)
  \label{eq5}
\end{split}
\end{equation}
\end{footnotesize}%
where $Z \in \mathbb{R}^{l \times d}$ is the input sequences with length $l$ and dimension $d$, and $Q_{i}, K_{i}, V_{i}\in \mathbb{R}^{l \times d/h}$ are the mapped queries, keys and values, respectively. $W^{Q}_{i}, W^{K}_{i}, W^{V}_{i} \in \mathbb{R}^{d \times d/h}$ and $W^{O} \in \mathbb{R}^{d \times d}$ are linear transformation matrices.

\subsection{Modified position-wise feed-forward network}
\label{sec:majhead}

A key issue for the transformer is how to exploit the order information in the speech sequence. Previous studies \cite{chen20l_interspeech,sperber2018self} have found that the positional encoding utilized in the original transformer is not suitable for dual-path networks. Inspired by the effectiveness of recurrent neural networks in tracking order information, a GRU layer is used as the replacement of the first fully connected layer in the feed-forward network to learn the location information \cite{sperber2018self}. The output of the multi-head self-attention is passed through the feed-forward network followed by residual and normalization layers to obtain the final output of the transformer.

\begin{footnotesize}
\begin{equation}
\begin{split}
   FFN(Mid) = ReLU(GRU(Mid))W_{1}+b_{1}
  \label{eq5}
\end{split}
\end{equation}
\end{footnotesize}
\begin{footnotesize}
\begin{equation}
\begin{split}
   Output=LayerNorm(Mid+FFN)
  \label{eq5}
\end{split}
\end{equation}
\end{footnotesize}%
where $FFN(\cdot)$ denotes the output of the position-wise feed-forward network, $W_{1} \in  \mathbb{R}^{d_{ff} \times d}$, $b_{1} \in  \mathbb{R}^{d}$, and $d_{ff}=4 \times d$.

\section{Proposed DPT-FSNet}
\label{sec:pagestyle}

In this section, we propose a frequency-domain dual-path transformer network for the SE task. As shown in Fig.1, our proposed model consists of an encoder, a dual-path transformer processing module (DPTPM), and a decoder.
\subsection{Encoder}
\label{ssec:subhead}

The encoder consists of a 1x1 convolutional layer and a dilated-dense block, where the dilated-dense block consists of four dilated convolutional layers. The input to the encoder is the complex spectrum $X \in \mathbb{R}^{2 \times T \times F}$ resulted from short-time  Fourier transform  (STFT), and the output is a high-dimensional representation $U \in \mathbb{R}^{C \times T \times F}$ with $C$ T-F spectral feature maps.
 
\subsection{Dual-path transformer processing module}
\label{sssec:subsubhead}

The DPTPM consists of two 1x1 convolutional layers, $B$ dual-path transformers (DPTs), and a gated 1x1 convolutional layer. Before the DPTs, we use a 1x1 convolutional layer to halve the channel dimension of the encoder output features to form a new 3-D tensor $D \in  \mathbb{R}^{C^{'} \times T \times F}$ ($C'=C/2$), and use $D$ as the input to the DPTs, as presented in Fig.1. Each DPT consists of an intra-transformer and an inter-transformer, where the intra-transformer models sub-band information and the inter-transformer models full-band information. Different from \cite{tang2021joint}, the DPT handles time and frequency paths alternatively instead of parallelly.

The intra-transformer processing block models the sub-band of the input features, which acts on the second dimension of $D$
\begin{footnotesize}
\begin{equation}
\begin{split}
   D^{intra}_{b} & = {IntraTransformer_{b}}[D^{inter}_{b-1}]\\ 
   & = [f^{intra}_{b}(D_{b-1}^{inter}[:,:,i], i=1,…, F)] 
  \label{eq5}
\end{split}
\end{equation}
\end{footnotesize}%
where $D^{intra}_{b}$ is the output of $IntraTransformer_{b}$, $f^{intra}_{b}$ is the mapping function defined by the transformer, $b=1,2 ..., B$ and $D^{inter}_{0} = D$ , $D_{b-1}^ {inter}[:,:,i] \in  \mathbb{R}^{C^{'} \times T}$ is the sequence defined by all the $T$ time step in the $i$-th sub-band. That is, the intra-transformer models the information of all time steps in each sub-band of the speech signal.

The inter-transformer processing block is used to summarize the information from each sub-band of the intra-transformer output to learn the global information of the speech signal, which acts on the last dimension of $D$

\begin{footnotesize}
\begin{equation}
\begin{split}
D^{inter}_{b} &= {InterTransformer_{b}}[D^{intra}_{b}] \\
              &= [f^{inter}_{b}(D_{b}^{intra}[:,j,:], j=1,…, T)]
  \label{eq5}
\end{split}
\end{equation}
\end{footnotesize}%
where $D^{inter}_{b}$ is the output of $InterTransformer_{b}$, $f^{inter}_{b}$ is the mapping function defined by the transformer, and $D_{b}^{intra}[:,j,:] \in  \mathbb{R}^{C^{'} \times F}$ is the sequence defined by the $j$-th time step in all $F$ sub-band. That is, the inter-transformer models the information of all sub-bands of the speech signal at each time step. With the intra-transformer, each time step in $D_{b}^{intra}$ contains all the information of the corresponding sub-band, which allows the inter-transformer to model the global (i.e., full-band) information of the speech signal.

The final output of the transformer $D^{inter}_{B}$ is passed through a 1x1 convolutional layer to double the channel dimension of the output feature and then through a gated convolutional layer to smooth the output value of the DPTPM. 

\subsection{Decoder}
\label{ssec:subhead}

The decoder consists of a 1x1 convolutional layer and a dilated-dense block, where the dilated-dense block is the same as in the encoder. The feature from the DPTPM output is passed through the decoder to obtain the estimated complex ratio mask \cite{williamson2015complex}. The enhanced complex spectrum is obtained by the element-wise multiplication between encoder's input and the mask, which is passed through the ISTFT to obtain the enhanced speech waveform.

\begin{table*}[th]
\caption{Comparison with other state-of-the-art systems on the VCTK+DEMAND dataset.}
\label{tab:word_styles}
\centering

\begin{tabular}{c c c c c c c c c}
\toprule  
 
Method & Domain & WB-PESQ  & STOI & CSIG & CBAK & COVL & Para. (M)  \\
\midrule

Noisy  & - & 1.97  & 0.91 & 3.34 & 2.44 & 2.63 & -  \\
\midrule
MetricGAN \cite{fu2019metricgan} & F & 2.86  & - & 3.99 & 3.18 & 3.42 & 1.90\\
TSTNN \cite{wang2021tstnn} & T & 2.96  & 0.95 & 4.33 & 3.53 & 3.67 & 0.92 \\
T-GSA \cite{kim2020t} & F & 3.06  & - & 4.18 & 3.59 & 3.62 & -\\ 
DEMUCS \cite{defossez20_interspeech} & T & 3.07  & 0.95 & 4.31 & 3.40 & 3.63 & 33.5 \\
SE-Conformer \cite{kim2021se} & T & 3.13  & 0.95 & 4.45 & 3.55 & 3.82 & - \\
Learnable Loss Mixup \cite{chang2021single} & F & 3.26  & - & 4.49 & 3.27 & 3.91 & 20.32\\
\midrule
\textbf{DPT-FSNet (Proposed)}  & F & \textbf{3.33}  & \textbf{0.96} & \textbf{4.58} & \textbf{3.72} & \textbf{4.00} & \textbf{0.88}\\

\bottomrule
\end{tabular}
\end{table*}

\subsection{Loss fuction}
\label{sssec:subsubhead}

In order to make full use of the time-domain waveform-level features and the  T-F domain spectrum features, our loss function combines both time-domain and T-F domain losses. The loss function is as follows:
\begin{footnotesize}
\begin{equation}
\begin{split}
   L = \alpha_{1} \times L_{audio} + \alpha_{2} \times L_{spectral}
  \label{eq1}
\end{split}
\end{equation}
\end{footnotesize}%
$L_{audio}$ is mean square error (MSE) loss:
\begin{footnotesize}
\begin{equation}
\begin{split}
   L_{audio} = \frac{1}{N}\sum_{i=0}^{N-1}(y_{i} - \tilde{y_{i}})^{2}
  \label{eq1}
\end{split}
\end{equation}
\end{footnotesize}%
where $y$ and $\tilde{y}$ are the sample of the clean speech and the enhanced speech, respectively. and $N$ denotes the number of samples in the waveform.
$L_{spectral}$ is L1 loss, which is defined as:
\begin{footnotesize}
\begin{equation}
\begin{split}
   L_{spectral} =& \frac{1}{TF}\sum_{t=0}^{T-1}\sum_{f=0}^{F-1}| (|Y_{r}(t,f)| - |\tilde{Y}_{r}(t,f)|) \\&+ (|Y_{i}(t,f)| - |\tilde{Y}_{i}(t,f)|) | 
  \label{eq1}
\end{split}
\end{equation}
\end{footnotesize}%
where $Y$ and $\tilde{Y}$ denote the spectrum of the clean speech and the spectrum of the enhanced speech, respectively. $r$ and $i$ are the real and imaginary parts of the complex spectrogram. $T$ and $F$ are the number of frames and the number of frequency bins, respectively

\section{Experiments}

\subsection{Dataset}

We use a small-scale and a large-scale dataset to evaluate the proposed model. For the small-scale dataset, we use the VCTK+ DEMAND dataset, which is widely used in SE research. This dataset contains pre-mixed noisy speech and its paired clean speech. The clean sets are selected from the VoiceBank corpus \cite{veaux2013voice}, where the training set contains 11,572 utterances from 28 speakers, and the test set contains 872 utterances from 2 speakers. For the noise set, the training set contains 40 different noise conditions with 10 types of noises (8 from DEMAND \cite{thiemann2013diverse} and 2 artificially generated) at SNRs of 0, 5, 10, and 15 dB. The test set contains 20 different noise conditions with 5 types of unseen noise from the DEMAND database at SNRs of 2.5, 7.5, 12.5, and 17.5 dB. All the utterances are downsampled to 16kHz. We use 4-second long segments. If an utterance is longer than 4 seconds, a random 4-second slice will be selected from that utterance.

For the large-scale dataset, we use the DNS dataset. The DNS dataset contains over 500 hours of clean clips from 2150 speakers and over 180 hours of noise clips from 150 classes. We simulate the noisy-clean pairs with dynamic mixing during training stage. Specifically, before the start of each training epoch, $75\%$ of the clean speeches are mixed with randomly selected room impulse responses (RIR) provided by \cite{reddy2021icassp}. By mixing the clean speech ($75\%$ of them are reverberant) and noise with a random SNR in between -5 and 20 dB, we generate the speech-noise mixtures. For evaluation, the DNS dataset has two non-blind test sets named $with\_reverb$ and $no\_reverb$, both of which contain 150 noisy-clean pairs.

\subsection{Experimental setup}

The window length and frame shift of STFT and ISTFT are 25ms and 6.25ms, respectively, and the FFT length is 512. The number of feature maps $C$ of the T-F spectrum is set to 64. All convolutional layers in the encoder and the decode are followed by layer normalization and parametric ReLU nonlinearity. Convolutional layers in the DPTPM are followed by parametric ReLU nonlinearity. The dense block consists of four dilated convolutional layers with dilated rate $d=2$. The number of input channels in the successive layers of the dense block increases linearly as $C$, $2C$, $3C$, $4C$, and the output after each convolution has $C$ channels. We use 4 stacked dual-path transformers, i.e., $B=4$ and $h=4$ parallel attention layers are employed. The hyperparameters $\alpha_{1}$, $\alpha_{2}$ in the Eq.(9) are set to 0.4 and 0.6, respectively.
In the training stage, we train the proposed model for 100 epochs. We use Adam \cite{kingma2014adam} as the optimizer and a gradient clipping with maximum L2-norm of 5 to avoid gradient explosion. A dynamic strategy \cite{vaswani2017attention} is used to adjust the learning rate during the training stage. 
\begin{footnotesize}
\begin{equation}
\label{eq6}
lr=\left\{
\begin{aligned}
&k_{1} \cdot d_{model}^{-0.5} \cdot n \cdot warmup^{-1.5}  , & n \leq warmup, \\
&k_{2} \cdot 0.98^{[epoch/2]}  , & n > warmup.
\end{aligned}
\right.
\end{equation}
\end{footnotesize}%
where $n$ is the number of steps, $d_{model}$ denotes the feature size of the input of the transformer, and $k_{1}$, $k_{2}$ are tunable scalars. In this paper, $k_{1}$, $k_{2}$, $d_{model}$, and $warmup$ are set to 0.2, $4e^{-4}$, 32, 4000, respectively.

\subsection{Evaluation metrics}

On both datasets, we use wide-band PESQ (dubbed WB-PESQ) \cite{rec2005p} and STOI \cite{taal2011algorithm} as evaluation metrics. WB-PESQ and STOI quantify the perceptual quality and the intelligibility of a speech signal, respectively. For the VCTK+DEMAND dataset, we also employ the three most commonly used metrics in the VCTK+DEMAND dataset, which are CSIG for signal distortion, CBAK for noise distortion evaluation, and COVL for overall quality evaluation \cite{hu2007evaluation}. CSIG, CBAK, and COVL are mean opinion score (MOS) predictors, with a score range from 1 to 5. For the DNS dataset, we also employ SI-SDR as evaluation metrics. Higher scores indicate better performance for all metrics.

\section{Experimental Results}

\subsection{Results on the VCTK+DEMAND dataset}

The proposed method is compared with other methods which also employ the same VCTK dataset. As shown in Table 1, our proposed model outperforms other transformer-based models such as TSTNN, T-GSA, SE-Conformer, and achieves state-of-the-art performance in terms of WB-PESQ, STOI, CSIG, CBAK, COVL with the least parameters.

\subsection{Ablation analysis}

The experimental results in the previous subsection demonstrate that our method improves the SE performance. To further validate the effectiveness of our method, we performed an ablation analysis. We designed four experiments labeled CED+Dual-path former, STFT+CED+BLSTM, STFT+CED+Sub-band former, STFT+CED+Full-sub former, which are abbreviated as \emph{exp.1}, \emph{exp.2}, \emph{exp.3}, \emph{exp.4} in the following. \emph{Exp.4} is our proposed method. The difference between \emph{exp.1} and \emph{exp.4} is that \emph{exp.1} take time-domain features as input, which replaces STFT/ISTFT with segmentation and overlap-add stage as in \cite{wang2021tstnn}. The difference between \emph{exp.3} and \emph{exp.4} is that the intra part and inter part of the dual-path transformer in \emph{exp.3} both model sub-band information as in Eq.7 while the inter part of the dual-path transformer in \emph{exp.4} model full-band information as in Eq.8. Same as \emph{exp.3}, the BLSTM in \emph{exp.2} only models sub-band information. For a fair comparison, the number of parameters and computational complexity of the models in the four experiments was essentially the same, and all use a window length of 25 ms and a frame shift of 6.25 ms to extract frames. Therefore, the system latency is also essentially the same for all four experiments. 

\begin{table}[th]
\caption{Ablation analysis results on the VCTK+DEMAND dataset}
\label{tab:word_styles}
\centering

\begin{tabular}{c c c c}
\toprule  
 
Method  & WB-PESQ  & STOI \\
\midrule
CED + Dual-path former & 2.97  & 0.95 \\
STFT + CED + BLSTM   & 3.05  & 0.95   \\
STFT + CED + Sub-band former  & 3.20  & 0.95  \\
STFT + CED + Full-sub former  & \textbf{3.33}  & \textbf{0.96}  \\

\bottomrule
\end{tabular}
\end{table}

By comparing \emph{exp.2} and \emph{exp.3}, we can see that the improved transformer performs better than BLTSM, which demonstrates the effectiveness of the improved transformer. Furthermore, \emph{exp.4} outperforms \emph{exp.3}, proving the advantages of fusing sub-band information and full-band information. In both \emph{exp.1} and \emph{exp.4}, the intra transformer and inter transformer in the dual-path transformer model local and global information, respectively. However, the evaluation results of \emph{exp.4} is much better than those of \emph{exp.1}, which proves that the frequency domain feature is more effective than the time domain feature for the dual-path transformer.

\subsection{Results on the DNS dataset}

Table 3 compares the metric scores of the proposed model with those of other architectures on the DNS dataset. We can see that our method outperforms the baseline. We noticed that compared with the full-band model, the proposed model has a more significant performance improvement on the reverberation data. We also see a similar trend in FullSubNet. The possible reason is that the full-band and sub-band fusion models include a sub-band model, and the sub-band model helps to model reverberation effects by focusing on the temporal evolution of the narrow-band spectrum.

\begin{table}[th]
\caption{Comparison with other state-of-the-art systems on the DNS $with\_reverb$ ($no\_reverb$) test sets.}
\label{tab:word_styles}
\centering

\begin{tabular}{c c c c}
\toprule  
 
Method & WB-PESQ & STOI (\%)  & SI-SDR (dB)\\
\midrule
Noisy                            & 1.82 (1.58)  & 86.62   (91.52)       & 9.03  (9.07)      \\
\midrule
NSNet~\cite{reddy2020interspeech}        & 2.37  (2.15)   & 90.43   (94.47)       & 14.72  (15.61)       \\
DTLN~\cite{westhausen20_interspeech}                 &   -     (-)               & 84.68    (94.76)      & 10.53  (16.34 )       \\
PoCoNet~\cite{isik20_interspeech}     & 2.83 (2.75)                       &      -     (-)         &       -     (-)                        \\
FullSubNet\cite{hao2021fullsubnet}      & 2.97 (2.78)      & 92.62 (96.11) & 15.75 (17.29)  \\
CTS-Net\cite{li2021two}      & 3.02 (2.94)      & 92.70 (96.66) & 15.58 (17.99)  \\
GaGNet\cite{li2021glance}      & - (3.17)      & - (97.13) & - (18.91)  \\
\midrule
\textbf{DPT-FSNet}   & \textbf{3.53} (\textbf{3.26}) & \textbf{95.23} (\textbf{97.68}) & \textbf{18.14} (\textbf{20.36}) \\
\bottomrule

\end{tabular}
\end{table}

\section{Conclusions}

In this paper, we propose a dual-path transformer-based full-band and sub-band fusion network for speech enhancement in the frequency domain. Inspired by the full-band and sub-band fusion models, we explore features that are more efficient for dual-path structures with the intra part in the dual-path transformer models the sub-band information, and the inter part models the full-band information. Experimental results on the Voice Bank + DEMAND dataset and DNS dataset show that the proposed method outperforms the current state of the art at a relatively small model size.

\bibliographystyle{IEEEbib}
 \patchcmd{\thebibliography}
  {\settowidth}
  {\setlength{\itemsep}{0pt plus 0.1pt}\settowidth}
  {}{}
\apptocmd{\thebibliography}
  {\footnotesize}
  {}{}   
\bibliography{ref}

\begin{thebibliography}{10}

\bibitem{loizou2013speech}
Philipos~C Loizou,
\newblock {\em Speech enhancement: theory and practice},
\newblock CRC press, 2013.

\bibitem{pandey2019new}
Ashutosh Pandey and DeLiang Wang,
\newblock ``A new framework for cnn-based speech enhancement in the time
  domain,''
\newblock {\em IEEE/ACM Transactions on Audio, Speech, and Language
  Processing}, vol. 27, no. 7, pp. 1179--1188, 2019.

\bibitem{defossez20_interspeech}
Alexandre Défossez, Gabriel Synnaeve, and Yossi Adi,
\newblock ``{Real Time Speech Enhancement in the Waveform Domain},''
\newblock in {\em Proc. Interspeech 2020}, 2020, pp. 3291--3295.

\bibitem{kim2021se}
Eesung Kim and Hyeji Seo,
\newblock ``Se-conformer: Time-domain speech enhancement using conformer,''
\newblock {\em Proc. Interspeech 2021}, pp. 2736--2740, 2021.

\bibitem{fu2019metricgan}
Szu-Wei Fu, Chien-Feng Liao, Yu~Tsao, and Shou-De Lin,
\newblock ``Metricgan: Generative adversarial networks based black-box metric
  scores optimization for speech enhancement,''
\newblock in {\em International Conference on Machine Learning}. PMLR, 2019,
  pp. 2031--2041.

\bibitem{kim2020t}
Jaeyoung Kim, Mostafa El-Khamy, and Jungwon Lee,
\newblock ``T-gsa: Transformer with gaussian-weighted self-attention for speech
  enhancement,''
\newblock in {\em ICASSP 2020-2020 IEEE International Conference on Acoustics,
  Speech and Signal Processing (ICASSP)}. IEEE, 2020, pp. 6649--6653.

\bibitem{hao2021fullsubnet}
Xiang Hao, Xiangdong Su, Radu Horaud, and Xiaofei Li,
\newblock ``Fullsubnet: A full-band and sub-band fusion model for real-time
  single-channel speech enhancement,''
\newblock in {\em ICASSP 2021-2021 IEEE International Conference on Acoustics,
  Speech and Signal Processing (ICASSP)}. IEEE, 2021, pp. 6633--6637.

\bibitem{li2021two}
Andong Li, Wenzhe Liu, Chengshi Zheng, Cunhang Fan, and Xiaodong Li,
\newblock ``Two heads are better than one: A two-stage complex spectral mapping
  approach for monaural speech enhancement,''
\newblock {\em IEEE/ACM Transactions on Audio, Speech, and Language
  Processing}, vol. 29, pp. 1829--1843, 2021.

\bibitem{liu20f_interspeech}
Haohe Liu, Lei Xie, Jian Wu, and Geng Yang,
\newblock ``{Channel-Wise Subband Input for Better Voice and Accompaniment
  Separation on High Resolution Music},''
\newblock in {\em Proc. Interspeech 2020}, 2020, pp. 1241--1245.

\bibitem{yang2021multi}
Geng Yang, Shan Yang, Kai Liu, Peng Fang, Wei Chen, and Lei Xie,
\newblock ``Multi-band melgan: Faster waveform generation for high-quality
  text-to-speech,''
\newblock in {\em 2021 IEEE Spoken Language Technology Workshop (SLT)}. IEEE,
  2021, pp. 492--498.

\bibitem{lv21_interspeech}
Shubo Lv, Yanxin Hu, Shimin Zhang, and Lei Xie,
\newblock ``{DCCRN+: Channel-Wise Subband DCCRN with SNR Estimation for Speech
  Enhancement},''
\newblock in {\em Proc. Interspeech 2021}, 2021, pp. 2816--2820.

\bibitem{takahashi2017multi}
Naoya Takahashi and Yuki Mitsufuji,
\newblock ``Multi-scale multi-band densenets for audio source separation,''
\newblock in {\em 2017 IEEE Workshop on Applications of Signal Processing to
  Audio and Acoustics (WASPAA)}. IEEE, 2017, pp. 21--25.

\bibitem{luo2020dual}
Yi~Luo, Zhuo Chen, and Takuya Yoshioka,
\newblock ``Dual-path rnn: efficient long sequence modeling for time-domain
  single-channel speech separation,''
\newblock in {\em ICASSP 2020-2020 IEEE International Conference on Acoustics,
  Speech and Signal Processing (ICASSP)}. IEEE, 2020, pp. 46--50.

\bibitem{chen20l_interspeech}
Jingjing Chen, Qirong Mao, and Dong Liu,
\newblock ``{Dual-Path Transformer Network: Direct Context-Aware Modeling for
  End-to-End Monaural Speech Separation},''
\newblock in {\em Proc. Interspeech 2020}, 2020, pp. 2642--2646.

\bibitem{zhang2021transmask}
Zining Zhang, Bingsheng He, and Zhenjie Zhang,
\newblock ``Transmask: A compact and fast speech separation model based on
  transformer,''
\newblock in {\em ICASSP 2021-2021 IEEE International Conference on Acoustics,
  Speech and Signal Processing (ICASSP)}. IEEE, 2021, pp. 5764--5768.

\bibitem{wang2021tstnn}
Kai Wang, Bengbeng He, and Wei-Ping Zhu,
\newblock ``Tstnn: Two-stage transformer based neural network for speech
  enhancement in the time domain,''
\newblock in {\em ICASSP 2021-2021 IEEE International Conference on Acoustics,
  Speech and Signal Processing (ICASSP)}. IEEE, 2021, pp. 7098--7102.

\bibitem{vaswani2017attention}
Ashish Vaswani, Noam Shazeer, Niki Parmar, Jakob Uszkoreit, Llion Jones,
  Aidan~N Gomez, {\L}ukasz Kaiser, and Illia Polosukhin,
\newblock ``Attention is all you need,''
\newblock in {\em Advances in neural information processing systems}, 2017, pp.
  5998--6008.

\bibitem{huang2017densely}
Gao Huang, Zhuang Liu, Laurens Van Der~Maaten, and Kilian~Q Weinberger,
\newblock ``Densely connected convolutional networks,''
\newblock in {\em Proceedings of the IEEE conference on computer vision and
  pattern recognition}, 2017, pp. 4700--4708.

\bibitem{valentini2016investigating}
Cassia Valentini-Botinhao, Xin Wang, Shinji Takaki, and Junichi Yamagishi,
\newblock ``Investigating rnn-based speech enhancement methods for noise-robust
  text-to-speech.,''
\newblock in {\em SSW}, 2016, pp. 146--152.

\bibitem{reddy2020interspeech}
Chandan~KA Reddy, Ebrahim Beyrami, Harishchandra Dubey, Vishak Gopal, Roger
  Cheng, Ross Cutler, Sergiy Matusevych, Robert Aichner, Ashkan Aazami,
  Sebastian Braun, et~al.,
\newblock ``The interspeech 2020 deep noise suppression challenge: Datasets,
  subjective speech quality and testing framework,''
\newblock {\em arXiv preprint arXiv:2001.08662}, 2020.

\bibitem{sperber2018self}
Matthias Sperber, Jan Niehues, Graham Neubig, Sebastian St{\"u}ker, and Alex
  Waibel,
\newblock ``Self-attentional acoustic models,''
\newblock {\em arXiv preprint arXiv:1803.09519}, 2018.

\bibitem{tang2021joint}
Chuanxin Tang, Chong Luo, Zhiyuan Zhao, Wenxuan Xie, and Wenjun Zeng,
\newblock ``Joint time-frequency and time domain learning for speech
  enhancement,''
\newblock in {\em Proceedings of the Twenty-Ninth International Conference on
  International Joint Conferences on Artificial Intelligence}, 2021, pp.
  3816--3822.

\bibitem{williamson2015complex}
Donald~S Williamson, Yuxuan Wang, and DeLiang Wang,
\newblock ``Complex ratio masking for monaural speech separation,''
\newblock {\em IEEE/ACM transactions on audio, speech, and language
  processing}, vol. 24, no. 3, pp. 483--492, 2015.

\bibitem{chang2021single}
Oscar Chang, Dung~N Tran, and Kazuhito Koishida,
\newblock ``Single-channel speech enhancement using learnable loss mixup,''
\newblock {\em Proc. Interspeech 2021}, pp. 2696--2700, 2021.

\bibitem{veaux2013voice}
Christophe Veaux, Junichi Yamagishi, and Simon King,
\newblock ``The voice bank corpus: Design, collection and data analysis of a
  large regional accent speech database,''
\newblock in {\em 2013 international conference oriental COCOSDA held jointly
  with 2013 conference on Asian spoken language research and evaluation
  (O-COCOSDA/CASLRE)}. IEEE, 2013, pp. 1--4.

\bibitem{thiemann2013diverse}
Joachim Thiemann, Nobutaka Ito, and Emmanuel Vincent,
\newblock ``The diverse environments multi-channel acoustic noise database
  (demand): A database of multichannel environmental noise recordings,''
\newblock in {\em Proceedings of Meetings on Acoustics ICA2013}. Acoustical
  Society of America, 2013, vol.~19, p. 035081.

\bibitem{reddy2021icassp}
Chandan~KA Reddy, Harishchandra Dubey, Vishak Gopal, Ross Cutler, Sebastian
  Braun, Hannes Gamper, Robert Aichner, and Sriram Srinivasan,
\newblock ``Icassp 2021 deep noise suppression challenge,''
\newblock in {\em ICASSP 2021-2021 IEEE International Conference on Acoustics,
  Speech and Signal Processing (ICASSP)}. IEEE, 2021, pp. 6623--6627.

\bibitem{kingma2014adam}
Diederik~P Kingma and Jimmy Ba,
\newblock ``Adam: A method for stochastic optimization,''
\newblock {\em arXiv preprint arXiv:1412.6980}, 2014.

\bibitem{rec2005p}
ITUT Rec,
\newblock ``P. 862.2: Wideband extension to recommendation p. 862 for the
  assessment of wideband telephone networks and speech codecs,''
\newblock {\em International Telecommunication Union, CH--Geneva}, 2005.

\bibitem{taal2011algorithm}
Cees~H Taal, Richard~C Hendriks, Richard Heusdens, and Jesper Jensen,
\newblock ``An algorithm for intelligibility prediction of time--frequency
  weighted noisy speech,''
\newblock {\em IEEE Transactions on Audio, Speech, and Language Processing},
  vol. 19, no. 7, pp. 2125--2136, 2011.

\bibitem{hu2007evaluation}
Yi~Hu and Philipos~C Loizou,
\newblock ``Evaluation of objective quality measures for speech enhancement,''
\newblock {\em IEEE Transactions on audio, speech, and language processing},
  vol. 16, no. 1, pp. 229--238, 2007.

\bibitem{westhausen20_interspeech}
Nils~L. Westhausen and Bernd~T. Meyer,
\newblock ``{Dual-Signal Transformation LSTM Network for Real-Time Noise
  Suppression},''
\newblock in {\em Proc. Interspeech 2020}, 2020, pp. 2477--2481.

\bibitem{isik20_interspeech}
Umut Isik, Ritwik Giri, Neerad Phansalkar, Jean-Marc Valin, Karim Helwani, and
  Arvindh Krishnaswamy,
\newblock ``{PoCoNet: Better Speech Enhancement with Frequency-Positional
  Embeddings, Semi-Supervised Conversational Data, and Biased Loss},''
\newblock in {\em Proc. Interspeech 2020}, 2020, pp. 2487--2491.

\bibitem{li2021glance}
Andong Li, Chengshi Zheng, Lu~Zhang, and Xiaodong Li,
\newblock ``Glance and gaze: A collaborative learning framework for
  single-channel speech enhancement,''
\newblock {\em arXiv preprint arXiv:2106.11789}, 2021.

\end{thebibliography}

\end{document}